\documentclass[a4paper,11pt]{article}
\usepackage{jinstpub} 
\usepackage{lineno}
\usepackage{enumitem}
\usepackage{graphicx}
\usepackage{subfigure}
\usepackage{caption}
\usepackage{soul}
\usepackage[sorting=none]{biblatex}
\title{\boldmath X-ray imaging with Micromegas detectors with optical readout}

\author[a]{A. Cools} \author[a]{, S. Aune} \author[b]{, F. Beau} \author[d]{, F.M. Brunbauer} \author[a]{, T. Benoit} \author[a]{, D. Desforge}\author[a]{, E.~Ferrer-Ribas}\author[a]{, A.~Kallitsopoulou} \author[b]{, C. Malgorn} \author[d]{, E. Oliveri} \author[a]{, T. Papaevangelou} \author[a]{, E.C. Pollacco} \author[d]{, L.~Ropelewski} \author[c]{, A. Sari}\author[e]{, F.J. Iguaz}
\affiliation[a]{IRFU, CEA, Université Paris-Saclay, F-91191 Gif-sur-Yvette, France}
\affiliation[b]{Institut Joliot, CEA, Université Paris-Saclay, F-91191 Gif-sur-Yvette, France}
\affiliation[c]{LIST, CEA, Université Paris-Saclay, F-91191 Gif-sur-Yvette, France}
\affiliation[d]{European Organization for Nuclear Research (CERN), 1211 Geneva 23, Switzerland}
\affiliation[e]{SOLEIL Synchrotron, L’Orme des Merisiers, Départementale 128, 91190 Saint-Aubin, France}

\emailAdd{antoine.cools@cea.fr}

\abstract{In the last years, optical readout of Micromegas gaseous detectors has been achieved by implementing a Micromegas detector on a glass anode coupled to a CMOS camera. Effective X-ray radiography was demonstrated using integrated imaging approach. High granularity values have been reached for low-energy X-rays from radioactive sources and X-ray generators.

Detector characterization with X-ray radiography has led to two applications: neutron imaging for non-destructive examination of highly gamma-ray emitting objects and beta imaging for the single cell activity tagging in the field of oncology drug studies.

First measurements investigating the achievable spatial resolution of the glass Micromegas detector at the SOLEIL synchrotron facility with a high-intensity and flat irradiation field will be shown in this article.
}


\arxivnumber{xxxx.xxxxx} 

\proceeding{7$^{\text{th}}$ International Conference on
Micro Pattern Gaseous
Detectors \\
11-16 December 2022\\
Weizmann Institute of Science, Rehovot, Israel}

\begin{document}
\flushbottom
\maketitle
\section{Introduction}
\label{sec:intro}
The optical readout of gaseous detectors is an emerging technology being already used in many different fields. Optical Time Projection Chambers (OTPC) are able to reconstruct 3D tracks in dark matter experiments~\cite{Baracchini_2020} and could also be used in the medical field for proton therapy beam monitoring~\cite{Meyer_2020}. Gaseous Electron Multipliers (GEM) detectors coupled with optical readout have also shown good performances regarding the gain and light yield for many different applications~\cite{Brunbauer:2632476}. Recent studies have shown that the Micromegas detector~\cite{GIOMATARIS1998239,BRUNBAUER2020163320} is well suited for imaging when it is implemented on a transparent anode and coupled with a CMOS camera. This detector obtained high spatial resolution under X-ray and neutron irradiation and was able to detect very low activity $\beta$-ray emitting samples~\cite{COOLS2023167910,JAMBON2022166332}.

One major strength of optical readout is the ability to integrate the light scintillated by the gas for real-time imaging. The device gives images with high contrast and granularity thanks to the Micromegas detector high gain and micromesh structure coupled with a CMOS camera of several mega pixels resolution.

A beam test at the SOLEIL synchrotron facility \cite{SOLEIL1,SOLEIL2} took place in November 2022 for characterization of the glass Micromegas detector. This test aimed to measure the spatial resolution and scintillation light response of the detector. This article describes the measurement of the Point Spread Function (PSF) of the detector for different configurations.

\section{Beam test at the SOLEIL synchrotron}
For a precise characterization of the optical readout Micromegas detector, it is essential to measure the impulse response of the detector. This is given by the PSF which describes the response of the detector to a point source. The SOLEIL synchrotron is a first-class solution, providing a very thin parallel beam of hard X-rays at very high flux. This test studies the origins of the blurring of the final image, namely the diffusion, the optical aberrations and the light reflection inside the detector. In this section, preliminary results from a test at the SOLEIL synchrotron are presented.
\subsection{Experimental setup}
This experiment was conducted at the SOLEIL Metrologie beam line \cite{SOLEIL1,SOLEIL2}, the main characteristics of which are described in this paragraph. The beam size is set at $5\times5$ µm². We assume that 5 µm is at a scale far lower than the intrinsic spatial resolution of the detector and the beam can be approximated to a point source. The photons are extracted from the storage ring of the synchrotron more than 20 m away from the detector and are expected to have parallel trajectories. The energy of the beam was set at 6 keV and the measured beam flux at the detector was $3.69\times10^5$ photons/s. The beam size was measured by a Basler camera \cite{Basler}. The camera, the power supply, the table position and the beam energy were controlled remotely from the control room.

The properties of the detector are depicted in details in \cite{COOLS2023167910} while the main features are presented here. The drift gap was set at 3 mm for this first measurement while a wide range of operating voltages was scanned. The Micromegas detector was built on a 5 mm quartz coated with 150 nm of Indium Thin Oxide (ITO). A standard mesh (18 µm wire diameter, 45 µm opening between wires) is attached to the crystal by the bulk process \cite{app11125362} with pillars of 500 µm diameter and 6 mm pitch. The gas mixture is composed of 90$\%$ of Argon and 10$\%$ of Carbon tetrafluoride $\mathrm{(CF_4)}$ which scintillation spectrum contains a wide emission band around 630 nm~\cite{Brunbauer:2632476}. The camera is an ORCA-Quest qCMOS imaging sensor from Hamamatsu \cite{Hamamatsu}, sensitive to single photons. The camera was coupled with a fast lens from Schneider group \cite{Schneider} of large aperture (f/0.95) for low light imaging applications. The detector and the camera are fixed on one solid plate to ensure the optical alignment and interconnected by a "light-tight" black cover made by 3D printing. The mesh was grounded while the cathode and the anode were at a negative and positive voltage respectively. The drift field ($\mathrm{E_{drift}}$) is defined between the cathode and the mesh and the amplification field ($\mathrm{E_{amp}}$) between the mesh and the anode. For every measurements, 100 frames of 1 s of beam signal were acquired.

\subsection{Analysis and results}
\begin{figure}[htbp]
\centering
\includegraphics[width=0.29\textwidth]{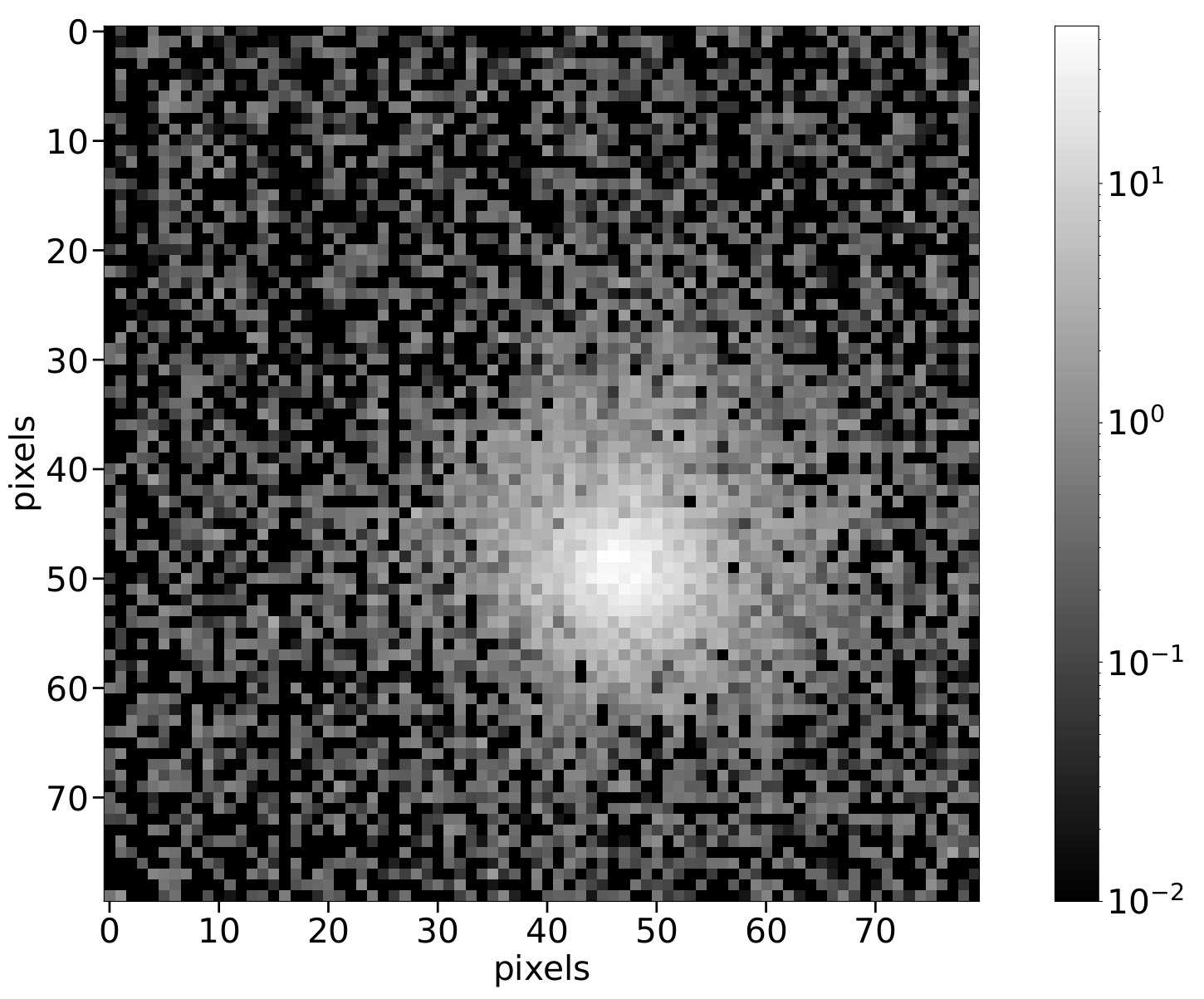}
\includegraphics[width=0.65\textwidth]{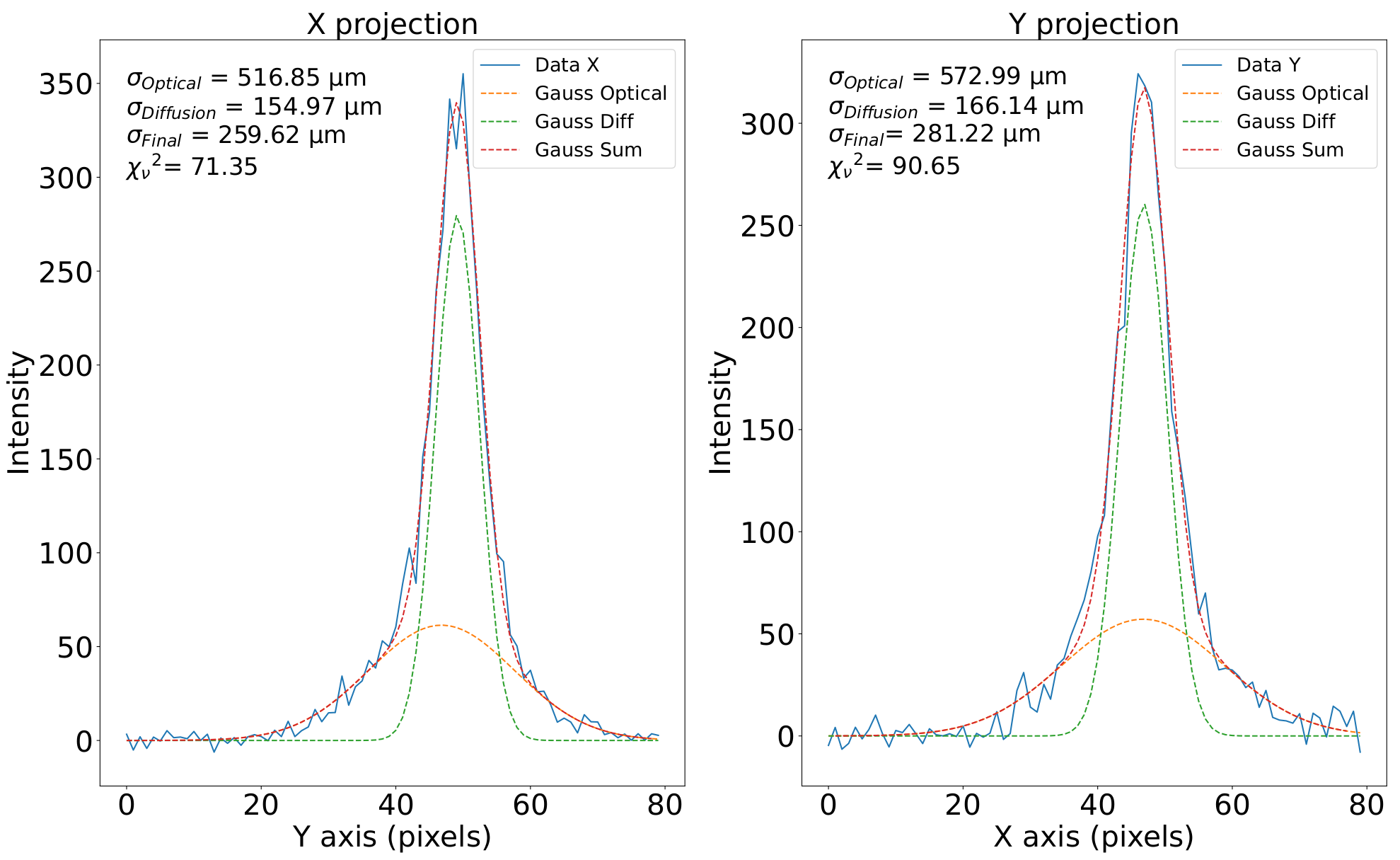}
\caption{2D intensity image (left) and 1D projections on horizontal and vertical axes (center, right) for $\mathrm{E_{amp}}=37~kV/cm$ and $\mathrm{E_{drift}}=330~V/cm$. }\label{fig:ii}
\end{figure}

Here we describe the image treatment processes and the different operations involved in the analysis. The background noise coming from the dark current and the readout noise of the camera constitutes the pedestal of the image. The pixels intensity of dark images correspond to 0 photons and varies from one pixel to another. For this reason, the mean value over 100 frames of every pixel i intensity ($\mu_i$) is computed and subsequently suppressed from the beam signal. Hence, the pedestal and most of the hot pixels are removed. In the follow-up analysis, the beam signal is characterized by the two 1D shapes instead of one 2D shape (Figure~\ref{fig:ii} left) by integrating the image over the horizontal and vertical axes (Figure~\ref{fig:ii} right). No intensity threshold is applied because the light spreading effects of low light intensity could be removed.

Figure~\ref{fig:ii} (right) represents the beam intensity profile projected along the horizontal and the vertical axes. Two main contributions to the light spreading have to be considered: the electron diffusion in the gas and the optical effects. The diffusion of charges by multiple collisions in the gas follows a Gaussian distribution with a standard deviation (STD) given by \begin{math}\sigma_x=\sqrt{2Dt}\end{math} after a time $t$, drifting through a distance $x$. The diffusion coefficient $D$ is a function of the charge velocity and strongly depends on the drift field. Thus, the diffusion increases as a function of \begin{math}\sqrt{x}\end{math} and is given in units of \begin{math}\mu m/\sqrt{cm}\end{math}. When a hard X-ray penetrates the gas volume, it can interact at any position in the drift gap, given a certain probability $\epsilon_x$ (absorption coefficient). Any photon interacting at a specific distance from the mesh implies a diffusion process of unique \begin{math}\sigma_x\end{math} that differs from other photons of a different position of interaction. Therefore, the contribution of the diffusion in Figure~\ref{fig:ii} (green cone) is defined by a sum of an infinite number of Gaussian curves with different \begin{math}\sigma_x\end{math} values. All the Gaussian distributions have the same mean value ($\mu_D$) because the X-ray beam is meant to be parallel to the drift field lines. Hence, the diffusion is modeled as one single Gaussian distribution of STD value $\sigma_D$ described in \eqref{eq:1a} where $\epsilon_x$ is the X-ray absorption coefficient at a distance $x$ from the cathode and $d$ is the drift gap thickness.

Nevertheless, optical effects such as optical aberrations or light reflections in the detector produce an additional contribution in the beam signal shape. Unlike optical TPCs where low luminosity events are captured, this detector aims to integrate high light intensity exposing the camera to third order aberrations. An important spreading pattern is brought from the optical aberration called Coma~\cite{optics}. Positive Coma causes a "comet-like" blur directed away from the optic axis which grows quadratically with the distance of the source from the optic axis. The lens has a large aperture ($f/0.95$) allowing photons that are far away from the optical axis to interact with the lens, increasing the aberration. As shown in Figure~\ref{fig:ii} (orange curve), if the X-ray beam does not hit the detector at the exact center in alignment with the camera, the beam signal shape becomes asymmetric. In other words, the Gaussian distribution associated to the optical effects of mean value $\mu_O$ is off-centered with the diffusion contribution. Furthermore, the inox micro-mesh being shinny, it produces reflections and adds some extra spread of the light. For the same reason, the cathode is made black to diminish reflections. In addition, the two 5 mm-thick quartz windows that light crosses before reaching the CMOS camera are sources of light reflection and refraction.

In order to calculate the uncertainty on the measurement of the beam size, the standard deviation of the sum of all the contributions is computed. The STD of {N} added up Gaussian shapes is given by \eqref{eq:1b}. Since $\mu_O\neq\mu_D$, the STD for {N=2} \eqref{eq:1c} contains an additional term involving the shift $(\mu_O-\mu_D)^2$ between the distributions. The STD and amplitude of the optical effects distribution are respectively $\sigma_O$ and $p_O$.
\begin{equation}\label{eq:1a} 
\begin{centering}
\sigma_D^2=\int_{0}^{d} \epsilon_x\sigma_x^2~dx
\end{centering}
\end{equation}

\begin{equation}\label{eq:1b} 
\begin{centering}
\sigma_{Final}^2=\sum_{i=1}^{N} p_i^2\sigma_i^2 + \sum_{i=1}^{N-1}\sum_{j=i+1}^{N} p_ip_j(\sigma_i^2+\sigma_j^2+(\mu_i-\mu_j)^2)
\end{centering}
\end{equation}

\begin{equation}\label{eq:1c}
\begin{centering}
\sigma_{Final}^2=(1-p_O)\sigma_D^2 + p_O\sigma_O^2 +p_O(1-p_O)(\sigma_D^2 + \sigma_O^2 +(\mu_D-\mu_O)^2)
\end{centering}
\end{equation}

\begin{figure}[htbp]
\centering
\includegraphics[width=0.49\textwidth]{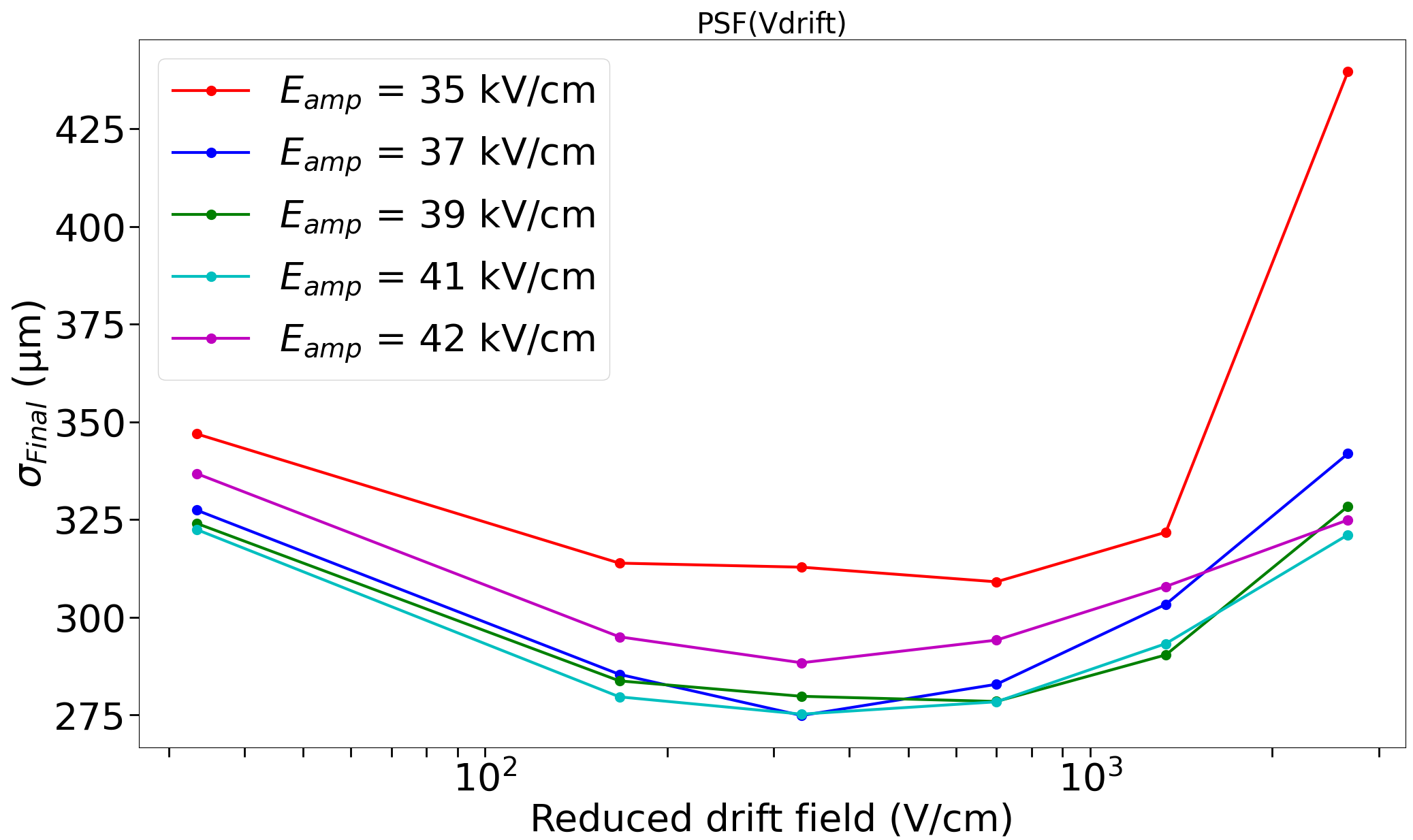}
\includegraphics[width=0.49\textwidth]{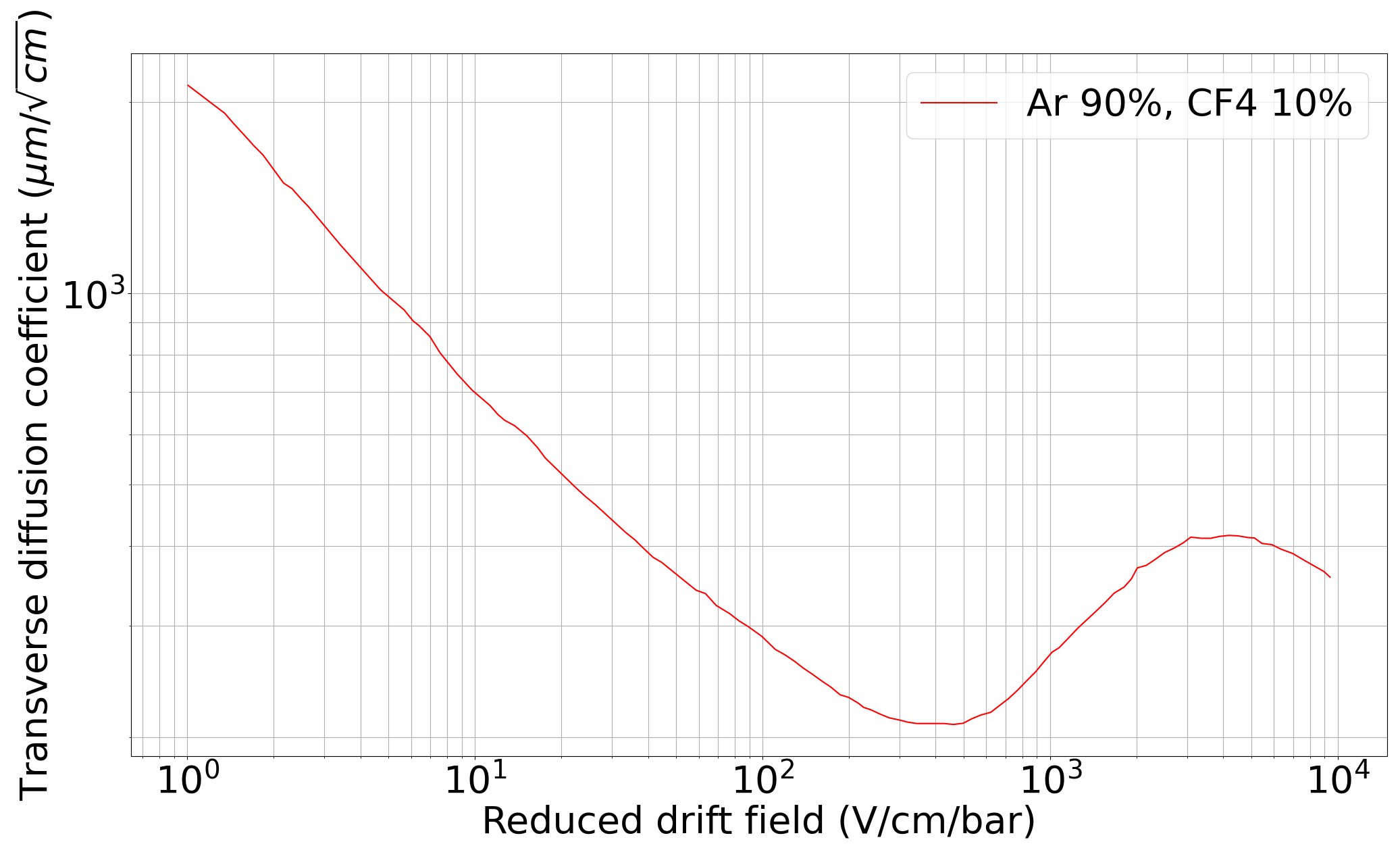}
\caption{STD of the beam signal as a function of the amplification and drift fields (left). Simulated diffusion coefficient for a gas mixture with $10\%$ of CF4 (right).}\label{fig:iii}
\end{figure}

The electric fields in the Micromegas detector are key parameters of the diffusion and amplification and directly affect the final signal standard deviation. Measurements were realized with five different amplification fields and six different drift fields in order to study the STD dependency on these fields. The minimum value of the STD was measured at $\mathrm{E_{drift}=330~V/cm}$ (Figure~\ref{fig:iii}, left) which corresponds to the minimum of the simulated diffusion coefficient (Figure~\ref{fig:iii}, right). At low amplification field (Figure~\ref{fig:iii}, left plot, red curve), the amount of light is small and the Signal to Noise Ratio is dominated by the readout noise, increasing the STD value. At high amplification field (Figure~\ref{fig:iii}, left plot, purple curve), the Micromegas detector enters the non-linear region and suffers from instabilities, increasing the light spreading.

\section{Conclusion}
A test run at the SOLEIL Metrologie beam line was performed for measuring of the Point Spread Function (PSF) of the optical Micromegas detector. The PSF of the device was computed at different electric field values allowing us to establish the dependency of the resolution on the diffusion and the amplification of the detector. Various sources of the light spreading have been identified as well as the effect of optical aberrations and the lens settings and will be studied further. Blackening of the mesh will be implemented in future set-ups for reflection investigation. The full analysis of the data is in progress and will allow us to exctract the response of the detector for different beam energies and sizes, detector positions and drift gaps. This will open the way to deconvolution of 2D images for high resolution imaging.

\acknowledgments

The authors acknowledge the financial support of the Cross-Disciplinary Program on Instrumentation and Detection (PTC-ID) of the French Alternative Energies and Atomic Energy Commission (CEA), of the «P2IO LabEx (ANR-10-LABX-0038)» in the framework "Investissements d’Avenir" (ANR-11-IDEX-0003-01) managed by the Agence Nationale de la Recherche (ANR), France and of the P2I Department of Paris-Saclay University. We acknowledge SOLEIL for provision of synchrotron radiation facilities (proposal number 99220033) and we would like to thank Pascal Mercere and Paulo Da Silva for assistance in using beamline METROLOGIE. The authors would like to thank Óscar Pérez and Luis Obis for their contribution to the data taking shifts during the beam test.

\printbibliography
\end{document}